\documentclass[a4paper]{jpconf}
\usepackage{graphicx}
\usepackage{amsmath, amsthm, amssymb, bm, braket}

\begin{document}
\title{Relativistic nuclear energy density functional approach to magnetic-dipole excitation} 

\author{Tomohiro Oishi$^{1}$, Goran Kru\u{z}i\'{c}$^{2}$, and Nils Paar$^{1}$}
\address{$^{1}$Department of Physics, Faculty of Science, University of Zagreb, Bijeni\v{c}ka c. 32, 10000 Zagreb, Croatia}
\address{$^{2}$Research department, Ericsson - Nikola Tesla, Krapinska 45, 10000, Zagreb, Croatia}

\ead{toishi@phy.hr}

\newcommand{\oprt}[1]{\hat{\mathcal{#1}}}
\newcommand{\abs}[1]{\left|  #1  \right|}
\newcommand{\slashed}[1] {\not\!{#1}} 

\def \beq{\begin{equation}}
\def \eeq{\end{equation}}
\def \beqa{\begin{eqnarray}}
\def \eeqa{\end{eqnarray}}

\def \adel{\tilde{l}} 

\begin{abstract}
Magnetic-dipole (M1) excitations of $^{18}$O and $^{42}$Ca nuclei are investigated 
within a relativistic nuclear energy density functional framework. 
In our last work \cite{2019OP}, 
these nuclei are found to have unique M1 excitation and its 
sum rule, because of their characteristic structure: 
the system consists of the shell-closure core plus two neutrons. 
For a more systematic investigation of the M1 mode, 
we have implemented a framework based on the relativistic nuclear energy density functional (RNEDF). 
For benchmark, we have performed the RNEDF calculations combined with the 
random-phase approximation (RPA). 
We evaluate the M1 excitation of $^{18}$O and 
$^{42}$Ca, whose sum-rule value (SRV) of the M1 transitions can be useful 
to test the computational implementation \cite{2019OP}. 
We also apply this RNEDF method to $^{208}$Pb, whose M1 property 
has been precisely measured \cite{1979Holt,1987Koehler,1988Laszewski,2016Birkhan}. 
Up to the level of the M1 sum rule, our result is in agreement 
with the experiments, except the discrepancy related with the quenching factors 
for $g$ coefficients. 
\end{abstract}

\section{Introduction}
The M1 excitation is one of the fundamental phenomena 
triggered by the electro-magnetic interactions with atomic nuclei. 
This is the leading mode to couple the unnatural-parity 
states, i.e. $J^{P}=1^+$ states. 
One can expect that, from the form of the M1 operator, 
its resonance can be useful for investigation of
spin-orbit level splitting, tensor-force effect, pairing correlations in medium \cite{2019OP}, etc. 
Noticeable collective motions, including scissors 
mode in deformed nuclei \cite{2010Heyde_M1_Rev}, can be also 
activated by the M1 excitation. 
Also, the analogy between the M1 and Gamow-Teller (GT) modes 
has attracted a special interest in recent studies \cite{2004Lang}. 
Indeed, zero component of the GT transition is almost identical 
to the isovector spin-M1 excitation. 
The GT resonance is expected as the dominant ingredient in 
neutrino-nucleus reactions in the energy scale of supernova, 
which can be a key to explain the origin of several elements. 
For an accurate evaluation of neutrino-nucleus reactions, 
certain theoretical framework, which can predict the GT as well as M1 
excitations throughout the nuclear chart, 
has been on a serious demand. 
See also Refs. \cite{2010Heyde_M1_Rev,1996Kneis_Exp_Rev,2008Pietralla_Rev} 
for more details on the M1 phenomena.

For an evaluation of the M1 mode without limitations on mass numbers, 
the mean field calculation based on the relativistic nuclear energy density 
functional (RNEDF) theory can be a suitable option \cite{1974Walecka,1977Boguta,1989Reinhard,2005Vret}. 
As an important feature of the M1 mode, its transition mainly 
occurs between the spin-orbit partners, e.g. 
$f_{7/2}\longrightarrow f_{5/2}$ for Ca isotopes. 
From the RNEDF effective Lagrangian, this spin-orbit 
splitting can be naturally concluded \cite{1989Reinhard,2005Vret}. 
On the other hand, 
for computations including unnatural-parity states, 
one should be careful for the residual interactions in RNEDF. 
These interactions do not contribute in the ground state (GS) with $J^{P}=0^{+}$, 
and thus, GS data cannot provide a reference to determine 
their model parameters. 
In order to optimize those parameters, one needs to shift focus 
to the measurable process, where the residual interaction plays 
an essential role. 
The M1 excitation is indeed a suitable reference for this purpose.

In this work, we develop a RNEDF-based framework to compute the 
collective M1 excitation. 
We adopt the CGS-Gauss system of units in this article.


\section{Formalism}
In the present study, the RNEDF framework has been employed to describe the nuclear ground state
properties within the relativistic mean field model at the Hartree level, and nuclear excitations are 
described using the relativistic random phase approximation (RPA). The respective formalism is derived 
from an effective Lagrangian density with four fermion contact interaction terms including
the isoscalar-scalar, isoscalar-vector, and isovector-vector channels \cite{1989Reinhard,2005Vret}. 
The effective Lagrangian contains the free-nucleon and density dependent interaction terms, coupling
of protons to the electromagnetic field, and the derivative term accounting for the leading 
effects of finite-range interactions necessary for a quantitative description of nuclear density 
distribution and radii. 
Detailed formalism and overview of the model calculations are given in Refs. \cite{2005Vret,2008Niksic,2014Niksic}.
In this work, we employ the point-coupling interaction DD-PC1 \cite{2008Niksic,2014Niksic}. 
The respective set of parameters for the RNEDF has been utilized in 
several applications, resulting in successful agreement with the experimental data on
nuclear ground state properties and excitations. 

In this work, we consider the collective M1 excitation 
of the $^{A}Z$ nucleus up to the one-body-operator 
level\footnote{We neglect the meson-exchange-current effect, for which one needs to consider the relevant multi-body terms.}: 
\beq
 \oprt{Q}_{\mu} ({\rm M1} ) \equiv  \sum_{k \in A=N+Z} \oprt{P}_{\mu}(k),
\eeq
where $\oprt{P}_{\mu}(k)$ with $\mu=0,\pm 1$ is the single-particle M1 operator for the $k$th nucleon. 
That is \cite{80Ring}, 
\beq
 \oprt{P}_{0}   = \mu_{\rm N} \sqrt{\frac{3}{4\pi}} \left( g_l \hat{l}_0 + g_s \hat{s}_0 \right),~~~
 \oprt{P}_{\pm} = (\mp) \mu_{\rm N} \sqrt{\frac{3}{4\pi}} \left( g_l \hat{l}_{\pm} + g_s \hat{s}_{\pm} \right), 
\eeq
where $\mu_{\rm N}$ is the nuclear magneton, 
$\hat{l}_0=\hat{l}_z$, 
$\hat{l}_{+}=(\hat{l}_x+i \hat{l}_y)/\sqrt{2}$, 
$\hat{l}_{-}=\hat{l}_{+}^{\dagger}$, 
and similarly defined for spin operators. 
Considering the different $g$ coefficients for protons and neutrons, 
$\oprt{Q}_{\mu}({\rm M1})$ reads 
\beq
 \oprt{Q}_{\mu} ({\rm M1} ) = 
 \mu_{\rm N} \sqrt{\frac{3}{4\pi}} \left[ 
   \sum_{i \in Z} \left(  g^{(p)}_s \hat{s}_{\mu}(i)   +g^{(p)}_l \hat{l}_{\mu}(i) \right)
 + \sum_{j \in N} \left(  g^{(n)}_s \hat{s}_{\mu}(j)   +g^{(n)}_l \hat{l}_{\mu}(j) \right) \right].
\eeq
Here $g_l=1~(0)$ and $g_s=5.586~(-3.826)$ for the proton (neutron) \cite{80Ring,70Eisenberg}. 
Note that, utilizing the isospin $\tau_0(k)=2t_0(k)=+1~(-1)$ for the $k$th proton (neutron), 
one can separate the collective M1 operator 
into the isoscalar (IS) and isovector (IV) terms. 
That is, 
\beqa
 \oprt{Q}_{\mu} ({\rm M1} ) &=& \oprt{Q}^{\rm IS}_{\mu} ({\rm M1} ) + \oprt{Q}^{\rm IV}_{\mu} ({\rm M1} ) \nonumber \\ 
 &=& \mu_{\rm N} \sqrt{\frac{3}{4\pi}} \sum_k \left[ \left( g^{\rm IS}_l\hat{l}_{\mu}(k) + g^{\rm IS}_s \hat{s}_{\mu}(k) \right) +\hat{\tau}_0(k) \left( g^{\rm IV}_l\hat{l}_{\mu}(k) + g^{\rm IV}_s\hat{s}_{\mu}(k)  \right)  \right],
\eeqa
where $g^{\rm IS}_l = g^{\rm IV}_l = \frac{1}{2}$, 
$g^{\rm IS}_s \equiv \frac{g^{(p)}_s+g^{(n)}_s}{2} =0.880$, and 
$g^{\rm IV}_s \equiv \frac{g^{(p)}_s-g^{(n)}_s}{2} = 4.706$. 
Notice that the IS spin-M1 response is often minor because of the cancellation of the $g$ coefficients. 

For the $g$ coefficients, so-called quenching factors have been utilized in 
M1 calculations \cite{1998VNC,2009Vesely,2010Nest}: 
$g^{\rm IS,IV}_{l,s}\longrightarrow \zeta g^{\rm IS,IV}_{l,s}$, 
where $\zeta$ often should be less than one for the agreement with 
experimental M1 data. 
This quenching effect is mainly from the the second-order configuration mixing, 
or equivalently, coupling with two-particle-two-hole states \cite{1988Takayanagi}. 
In this article, however, we fix $\zeta=1$, except the 
case with special mentioning.

The M1 excitation strength is evaluated as 
\beq
  B_{\rm M1}(E_{\gamma}) = \sum_{\mu =0,\pm 1} \abs{\Braket{f | \oprt{Q}_{\mu}({\rm M1} ) |  i} }^2,
\eeq
where $E_{\gamma}=E_f -E_i$ is the excitation energy. 
For this evaluation, we utilize the random-phase approximation (RPA). 
Namely, the same procedure has been employed as in Refs. \cite{2003Paar,2005Niksic} , but 
in the present analysis the relativistic point coupling interaction is used, with the parameterization
DD-PC1. In the transition matrix elements, the magnetic operator $\oprt{Q}_{\mu}({\rm M1} )$ is used.
Then $B_{\rm M1}(E_{\gamma})$ is evaluated for the excitation from 
the $0^{+}$ ground state (GS), $\ket{i}$, to the $1^{+}$ excited state, $\ket{f}$. 
We assume the spherical symmetry in this work. 

\begin{table}[tb]
\caption{Sum-rule values of M1 ($S_{\rm M1}$) obtained in this work. 
The unit is $\mu_{\rm N}^2$. 
The corresponding analytic result in Ref. \cite{2019OP} with the three-body model 
(3BM) is also shown for comparison. }
\label{table:app_1_1}
\begin{center}
\begin{tabular}{lcc}
\br
                            &This work    &Ref. \cite{2019OP}   \\
\mr
  Method                    &RNEDF (DD-PC1)        &3BM  \\
  $S_{\rm M1}$ for $^{18}$O    &$2.73$                &$2.79$ \\
  $S_{\rm M1}$ for $^{42}$Ca   &$2.91$                &$2.99$ \\
  $S_{\rm M1}$ for $^{208}$Pb  &$52.86$\phantom{$9$}  &$-$~~ \\
\br
\end{tabular}
\end{center}
\end{table}

\section{Result and Discussion}
\subsection{No-pairing sum rule in {\rm $^{18}$O} and {\rm $^{42}$Ca} }
In order to check the numerical implementation, 
sum rules of the excitation strength often provide a useful guidance. 
For the M1 mode, we can refer to one case-limited but available 
version of its sum rule in Ref. \cite{2019OP}. 
There, the non-energy-weighted sum-rule value (SRV) of the M1 
excitation was evaluated for some specific systems, which 
consist of the shell-closure core and two valence 
neutrons or protons. 
Those systems include, e.g. $^{18}$O and $^{42}$Ca. 
An advantage of that sum rule is that, when the pairing 
correlation between the valence nucleons 
is neglected, the SRV is determined analytically 
for the corresponding system of interest. 

\begin{figure}[tb]
  \begin{minipage}{17pc}
    \includegraphics[width=17pc]{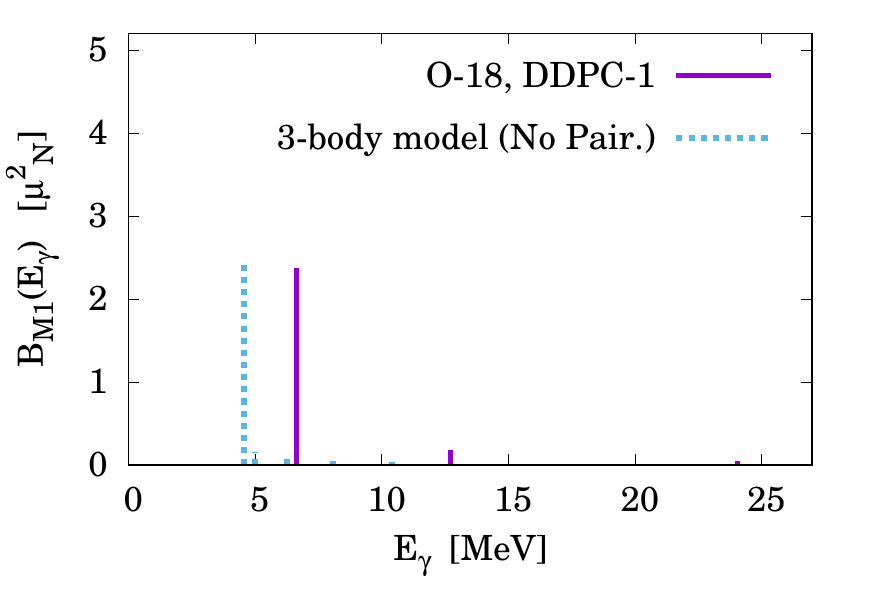}
    \end{minipage}\hspace{2pc}%
    \begin{minipage}{17pc}
    \includegraphics[width=17pc]{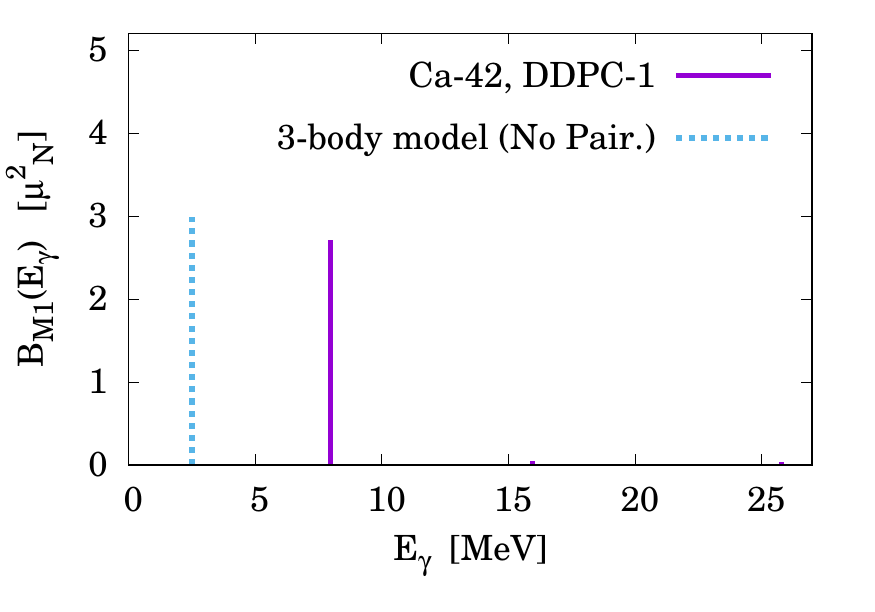}
  \end{minipage} 
\caption{(Left panel) The M1 strength $B_{\rm M1}(E_{\gamma})$ of $^{18}$O for DD-PC1 interaction. 
The result with the three-body model in Ref. \cite{2019OP} is also plotted 
for comparison. 
(Right panel) Same plot but for $^{42}$Ca. }
  \label{fig:1221_2018}
\end{figure}

The numerical SRV is determined as 
\beq
  S_{\rm M1} \equiv \int dE_{\gamma} B_{\rm M1}(E_{\gamma}), 
\eeq
where $E_{\gamma}$ is the excitation energy. 
Our results for the M1 SRV are shown in Table \ref{table:app_1_1} and in Figure \ref{fig:1221_2018}
 for $^{18}$O and $^{42}$Ca. In our RNEDF calculations, the pairing energy is neglected, in order to 
keep consistency between the no-pairing result in Ref. \cite{2019OP}. 
 For comparison, the results of the three-body model are also shown.
We take strength values up to $60$ MeV into account for this SRV. 
The actual $B_{\rm M1}(E_{\gamma})$ distribution is plotted in Figure \ref{fig:1221_2018}. 
There is one significant peak each in $^{18}$O and $^{42}$Ca. 
From the configuration results in RNEDF plus RPA calculations, 
it is found that these peaks are attributable to 
the neutron transitions of $d_{5/2}\longrightarrow d_{3/2}$ and $f_{7/2}\longrightarrow f_{5/2}$ 
in $^{18}$O and $^{42}$Ca, respectively. 
Namely, only the two valence neutrons are active for M1 transitions in these systems.

As displayed in Table \ref{table:app_1_1}, our 
SRV is obtained consistently to that in Ref. \cite{2019OP}. 
Thus, our implementation can be valid in the 
level of no-pairing sum rule. 
Note that, for $^{18}$O or $^{42}$Ca, the accurate measurement of M1 
strength has not been achieved yet. 

The excitation energy (position of the peak) of $B_{\rm M1}(E_{\gamma})$ 
shows an unnegligible difference between Ref. \cite{2019OP} and this work. 
This problem can be independent of the no-pairing SRV, but should be 
related with the spin-orbit splitting energies from the two models. 
We would like to remind that, in Ref. \cite{2019OP}, the $B_{\rm M1}(E_{\gamma})$ 
distribution is shown to be sensitive to the choice of the pairing model. 
Also, in the RNEDF side, 
the M1-excitation energy is expected to depend on the 
residual interactions, especially pseudo-vector interaction. 
For more precise discussions, one needs to optimize these 
model parameters with respect to the experimental data. 
More details will be given in forthcoming publication \cite{2019Kruzic}.

\begin{figure}[b]
  \includegraphics[width=17pc]{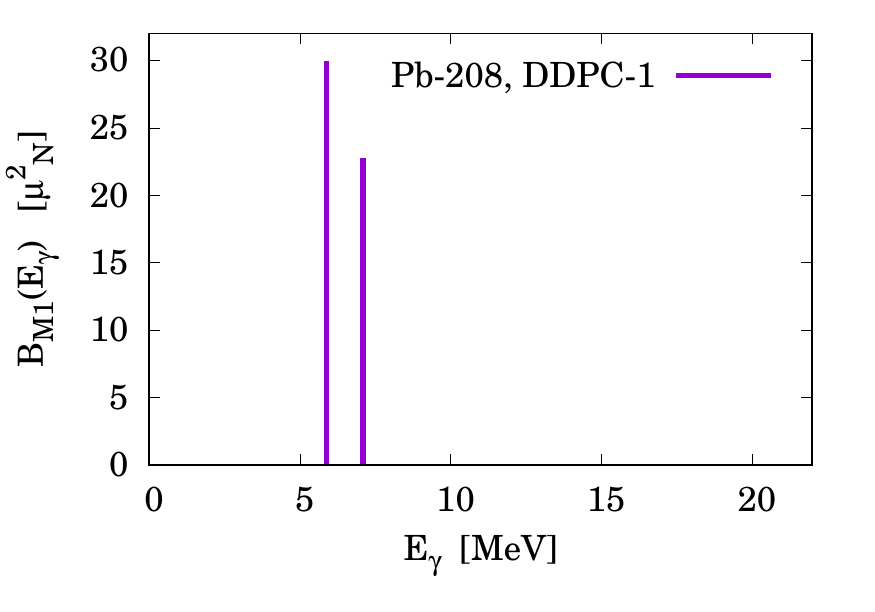}\hspace{2pc}%
  \begin{minipage}[b]{17pc}\caption{Same as Figure \ref{fig:1221_2018} but for $^{208}$Pb nucleus. } \label{fig:Pb208}
  \end{minipage}
\end{figure}

\subsection{Result in {\rm $^{208}$Pb}}
Following the consistency test of no-pairing SRV, 
we next apply our RNEDF framework to the $^{208}$Pb nucleus. 
This nucleus is one of the most precisely measured 
systems with respect to the M1 excitations. 
Its mean-excitation energy is measured as 
$E_{\gamma} = 7.3$ MeV \cite{1979Holt,1987Koehler,1988Laszewski,2016Birkhan}. 
The total SRV of M1 is also evaluated as 
$S_{\rm M1}= 15-20~\mu^2_{\rm N}$ \cite{1988Laszewski,2016Birkhan}.

In Figure \ref{fig:Pb208}, the calculated $B_{\rm M1}$ distribution is plotted. 
The mean-excitation energy is in a good agreement with experiments. 
As a remarkable difference from $^{18}$O or $^{42}$Ca, 
the M1 strength shows two peaks. 
This result is consistent to the two-peak structure 
in the experimental data \cite{1988Laszewski,2016Birkhan}. 
The origin of this two-peak distribution is simple: 
in $^{208}$Pb, both the valence protons in the $0h_{11/2}$ orbit 
and neutrons in the $0i_{13/2}$ orbit 
can be available for the M1 transition. 
Note also that the other spin-orbit-partner levels are fully occupied, 
and thus, cannot be active.


The SRV is obtained as $S_{\rm M1} =52.86~\mu^2_{\rm N}$ from our calculation. 
This value indeed overshoots the experimental result \cite{1988Laszewski,2016Birkhan}. 
Here we should mention that, in our calculations, 
the quenching factor $\zeta$ on the $g$ coefficient has been fixed as one. 
In some literature \cite{1998VNC,2009Vesely,2010Nest}, 
however, it has been suggested 
that $\zeta \simeq 0.6-0.7$ is necessary for consistency 
with experimental data. 
Notice that $S_{\rm M1}$ as well as $B_{\rm M1}(E_{\gamma})$ should be reduced by $\zeta^2$. 
This procedure then concludes the result $S_{\rm M1} \simeq 20~\mu^2_{\rm N}$, which is in
a reasonable agreement with the experimental data. 
Note that the quenching factor does not change the excitation energy $E_{\gamma}$, 
which may be shifted only by changing the RNEDF parameters.

\section{Summary and Outlook}
We have developed the RNEDF framework using the RPA in order
to investigate the properties of M1 excitations in nuclei. 
In the benchmark test for $^{18}$O and $^{42}$Ca, our result consistently 
reproduces the no-pairing SRV \cite{2019OP}. 
We have also investigated the $^{208}$Pb, and found that, 
except for the quenching effect, our results are consistent 
to the experimental energy and the SRV. We note that in the forthcoming
study the role of the pseudo-vector interaction terms in the residual RPA
interaction will be studied in more details. In addition, the pairing effects 
on M1 transitions will be studied using the relativistic quasiparticle random
phase approximation.

\ack
We sincerely thank A. Tamii for suggestions from the experimental side.
This work is supported by  the QuantiXLie Centre of Excellence, 
a project co financed by the Croatian Government and European Union through 
the European Regional Development Fund, the Competitiveness and Cohesion 
Operational Programme (KK.01.1.1.01).

\section*{References}
\providecommand{\newblock}{}


\end{document}